\documentclass[fleqn,twoside]{article}
\usepackage{espcrc2}
\usepackage{epsfig}

\usepackage{graphicx}
\usepackage[figuresright]{rotating}

\newcommand{\AmS}{{\protect\the\textfont2
  A\kern-.1667em\lower.5ex\hbox{M}\kern-.125emS}}

\newcommand{\half}{{\textstyle{\frac{1}{2}}}}

\newcommand{\bgeq}{\begin{equation}}
\newcommand{\bgeqa}{\begin{eqnarray}}
\newcommand{\edeq}{\end{equation}}
\newcommand{\edeqa}{\end{eqnarray}}

\newcommand{\ainv}{a^{-1}}

\newcommand{\aive}{a\"\i{}ve }
\newcommand{\msbar}{\overline{\rm MS}}

\newcommand{\PRD}[1]{Phys.\ Rev.\ \textbf{D{}#1}}

\title{
On the strange quark mass with improved staggered quarks
}

\author{J.~Hein\address{School of Physics, The University of Edinburgh, 
         Edinburgh EH9 3JZ, Scotland, UK},
        C.~Davies\address{Department of Physics \& Astronomy,
        The University of Glasgow, Glasgow G12 8QQ, Scotland, UK},
        G.~P.~Lepage\address{Newman Laboratory of Nuclear Studies, 
        Cornell University, Ithaca, NY 14850, USA},
        Q.~Mason\addressmark{}
        and
        H.~Trottier\address{Physics Department, Simon Fraser University, Burnaby,
        B.C., Canada V5A 1S6} (HPQCD and UKQCD collaboration)}

\begin{document}

\begin{abstract}
We present results on the sum of the masses of light and strange quark
using improved staggered quarks. Our calculation uses 2+1 flavours of
dynamical quarks. The effects of the dynamical quarks are clearly
visible. 
\end{abstract}

\maketitle

\noindent\hspace*{-1.9mm}
\raisebox{8cm}[0ex][0ex]{
{\normalsize
\parbox{4cm}{
\textbf{\textsf{Edinburgh 2002/15}}\\
\textbf{\textsf{GUTPA/02/09/01}}\\
\textbf{\textsf{hep-lat/0209077}}}}
}\vspace*{-5ex}

\section{INTRODUCTION}

\subsection{Motivation}
Simulations of hardronic matter with realistic sea quark effects form
a long standing problem in lattice gauge theory. To achieve this goal
on the upcoming generation of computers, improved staggered quarks are
most promising.
Compared to fermion formulations based on the Wilson
action, staggered quarks have fewer variables and are faster to
simulate for small quark masses.

N\aive staggered quarks suffer from large flavour changing strong
interactions, which can be suppressed by the use of fat links
\cite{blum}. This action can be further improved to
$\mathcal{O}(\alpha_sa^2,a^4)$ \cite{flsym}, the most precise to
date, see \cite{qtalk} for recent developments. These actions have
proven to be successful to reduce the splittings in the pion spectrum
\cite{milcvar} and the large renormalisations \cite{se_pap,op_pap}, as
observed for n\aive staggered quarks.  Due to the advantages of
improved staggered quarks it has been possible to generate data sets
with $2+1$ flavours of dynamical sea quarks with a low
$m_\pi/m_\rho$-ratio \cite{milcdata,doug_poster}.  An alternative
proposal for the improvement of staggered quarks is the hypercubic
action \cite{HK_hyp}.

Staggered quarks have a residual ``chiral'' symmetry, which protects
against additive mass renormalisations. The point of vanishing
quark mass is exactly known beforehand. This removes the dependency on
chiral extrapolations in the determination of the strange quark
mass.

\subsection{Simulation data}
For our calculation, we use the simulation results of the MILC
collaboration \cite{milcdata}. They use a fermionic action corrected
to $\mathcal{O}(\alpha_sa^2,a^4)$ \cite{flsym}, which is referred to as
ASQTAD. There are results for  0, 2 and 2+1 flavours of dynamical
fermions. Their lightest quark mass gives a low $m_\pi/m_\rho \approx
0.37$. The simulation parameters of the configurations have been
tuned to keep the lattice spacing fixed in units of the gluonic observable
$r_1$.

\section{LATTICE SPACING}

\subsection{Upsilon spectrum}
$\Upsilon$-spectroscopy on the configurations of
\cite{milcdata} is discussed in \cite{alantalk}. For the data set with
$n_f=2+1$ and the smallest $m_\pi/m_\rho$ ratio one gets $\ainv =
1.6 \mbox{ GeV}$.  In the quenched approximation the inverse lattice spacing
is substantially larger.

\subsection{Kaon spectrum}
In real world, the difference $m^2_{\rm V}-m^2_{\rm PS}$ between the
square of the mass of the vector and the pseudo-scalar meson is almost
independent from the masses of the valence quarks. In principle this
should yield a good method for determining the lattice spacing. Vector
mesons are unstable however. Their decays will eventually be seen in
dynamical lattice results, but currently it is unclear what effect
this has on the measured masses.

In the quenched approximation,  $m^2_{\rm V}-m^2_{\rm PS}$
is much more sensitive to the valence quark masses
than the real world..
For data sets with dynamical fermions on the other hand 
the splittings $m^2_{\rho}-m^2_{\pi}$,
$m^2_{K^*}-m^2_{K}$ and $m^2_{\phi}-m^2_{\eta_s}$ differ by only $\approx
2\%$. 
On the $n_f=2+1$ sets one gets 
$
\ainv = 1.4 \mbox{ GeV}
$ from $m^2_{K^*}-m^2_{K}$.
For the quenched data this procedure gives an
inverse lattice spacing which is about 100~MeV smaller.

\subsection{Gluonic scale}
The gluonic observable $r_1$ is determined in a similar way to
$r_0$, but gives the point where $r^2F(r)=1$. Since the configurations
have been generated for fixed $r_1$, it is natural to use $r_1$ as an
alternative to set the scale. MILC estimated
$r_1=0.35$~fm, using $r_0=0.5$~fm. Neither of these quantities is
easily related to a physical observable, however, and so there must be
considerable uncertainty in their physical value. Taking $r_1 =
0.35$~fm gives
$
\ainv = 1.5\mbox{ GeV}
$
for all configurations.

\section{QUARK MASS}

\subsection{Determination of the bare quark mass}
\begin{figure}[t]
\vspace*{2mm}
\centerline{\epsfig{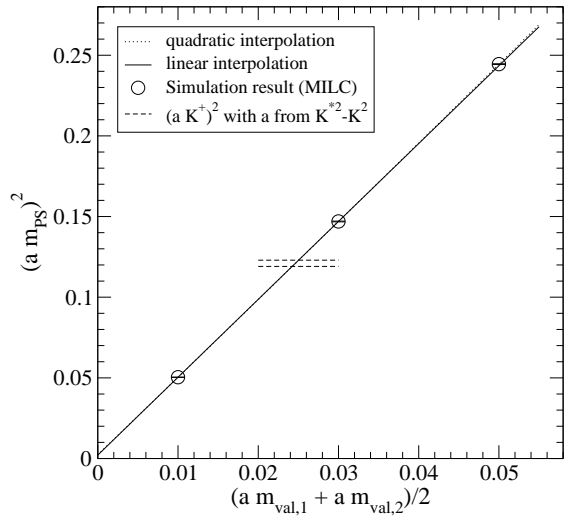}}
\vspace*{-9mm}
\caption{\label{mps_int_plot} 
Interpolation of the pseudoscalar mass to the Kaon
}
\end{figure}

We determine the sum of the bare masses of the light and strange quark
by interpolation of the pseudo-scalar meson mass to the physical $K$
with the ansatz
\bgeq
(am_{\rm PS})^2 = a + b\: am_{\rm av} + c (am_{\rm av})^2\,.
\edeq
With $m_{\rm av}:=\half(m_{\rm 1}+m_{\rm 2})$ we denote the
average of the valence masses. This is shown in
figure~\ref{mps_int_plot} for the most chiral dynamical data set.
The data points represent the $\pi$, $K$ and $\eta_s$. The difference
between a quadratic and linear interpolation is just visible in the
upper right corner of the figure. At the physical $K$ this difference is
completely irrelevant.

At this point the advantage of not having additive mass renormalisations
for staggered quarks pays back. No simulation is needed to
determine the point of vanishing quark mass.

\subsection{Renormalisation and matching}
The renormalisation and matching to the $\msbar$-scheme is done in a
three step procedure. We use the quark self-energy to convert from the
lattice mass $m_0$ to the pole mass
\cite{se_pap}. In a second step we convert to the $\msbar$-mass at
the scale of the
pole mass and finally run to the scale $\mu=2$~GeV. At 1-loop level we
get 
\bgeqa
{m_{\msbar}}
&=&\left[1+\alpha_s\!\left(b-\frac4{3\pi}-
\frac2\pi\log(a\mu)\right)\right]m_0\nonumber \\
&=:& \left[1 + \alpha_s Z_0^{(2)} \right]\,m_0\,.
\edeqa
Here
$
b:= \delta m_0^{(2)}/m_0 - 2/\pi\,\log(am_0)
$.
This is the 1-loop contribution to the mass shift between the bare
lattice mass and the pole mass with the logarithmic divergence taken
out. For small values of $am_0$ this becomes a constant \cite{se_pap}.
For $n_f=2+1$ we obtain
\bgeq
-0.12 < Z_0^{(2)} < -0.04\,,
\edeq
depending on the quantity used to fix the lattice spacing. It is
interesting to note, for n\aive staggered quarks, that is without fat
links to suppress the flavour changes, $Z_0^{(2)}\approx 3$.

For $\alpha_s$ we use the average of $\alpha_P$ for the scales $aq=1$
and 2, determined on these configs \cite{cthd_proc}. The difference
between the two is included into the error. Preliminary calculations
of the modified BLM scale $aq^*$ \cite{scale_setting} give values
lying between 1 and 2. Because of the small $Z_0^{(2)}$, variations in
$q$ have only very small effects.

\begin{figure}[t]
\vspace*{2mm}
\centerline{\epsfig{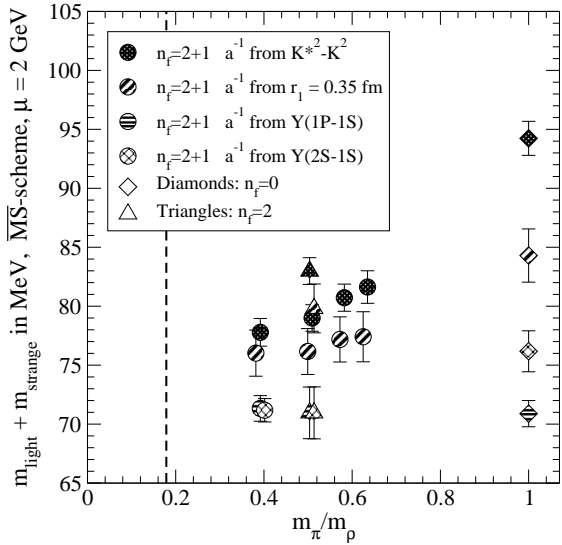}}
\vspace*{-9mm}
\caption{\label{mren_plot} Sum of light and strange quark mass in the
$\msbar$-scheme at a scale $\mu=2$~GeV vs the $m_\pi/m_\rho$-ratio of
the lightest sea quarks.}
\end{figure}

Our final result for the sum of the masses of the light and the
strange quark in the $\msbar$-scheme is presented in
figure~\ref{mren_plot}. Different shapes of the symbols
refer to the numbers of sea quarks. The different shadings of
the symbols indicate the determination of the lattice spacing. The
dashed line corresponds to the real world $m_\pi/m_\rho$-ratio.
The error bars encompass statistical uncertainties, the uncertainties
of the determination of $a$ and the uncertainty of the scale setting
of $\alpha_s$. One needs to add an uncertainty for the 2-loop
effects. At the present time we estimate this as
$\mathcal{O}(\alpha_s^2) \approx 20\%$ for $n_f=2+1$ and smaller
otherwise.

In the quenched case, the results show a wide spread,
depending on how $a$ is determined. This spread is clearly reduced for
$n_f=2$ and even smaller for $n_f=2+1$. There might even be a trend
for a decrease with decreasing $m_\pi/m_\rho$. In
table~\ref{restab} we give preliminary results from the most chiral
point of the corresponding data set.
\begin{table}[t]
\caption{\label{restab}Preliminary results on $m_{\rm
l}+m_{\rm s}$ in $\msbar$-scheme.}
\begin{tabular}{ccc}
\hline
$n_f$    & $a(K^{*2}-K^2)$   & $a(\Upsilon(1P)-\Upsilon(1S))$ \\\hline
0        & 94(9) MeV         & 71(7) MeV    \\
2        & 83(11) MeV        & 71(10) MeV   \\
2+1      & 78(14) MeV        & 71(13) MeV   \\ \hline
\end{tabular}
\end{table}

\section{SUMMARY}
We presented the first ever calculation of the light quark masses
using two flavours of dynamical light quarks and one flavour of
dynamical strange. This has been possible due to the use of improved
staggered quarks. Due to this improvement, the renormalisations are
small and well controlled. The dynamical quarks clearly reduce the
dependency on the quantity used to determine the lattice
spacing. Compared to other lattice calculations and sum rule
calculations, see \cite{msref} resp.\ \cite{mssum} for review,
our results favour a low value of
the strange quark mass.

\subsection*{Acknowledgement}
This work is supported by the EU~IHP programme, NSF, NSERC and PPARC.

\end{document}